\documentstyle[epsf,prd,aps,floats]{revtex}
\input{psfig.tex}
\newcommand {\be} {\begin{equation}}
\newcommand {\ee} {\end{equation}}

\begin{document}
\twocolumn[\hsize\textwidth\columnwidth\hsize\csname
@twocolumnfalse\endcsname

\title{The Behaviour of Varying-Alpha Cosmologies}
\author{John D. Barrow$^a$, H\aa vard Bunes Sandvik$^b$, and Jo\~{a}o Magueijo$^b$ \\
%EndAName
$^a$DAMTP, Centre for Mathematical Sciences, \\
Cambridge University, Wilberforce Road, \\
Cambridge CB3 OWA, U.K.\\
$^b$Theoretical Physics, The Blackett Laboratory, \\
$\ $Imperial College, Prince Consort Road, \\
London SW7 2BZ, U.K.}
%\date{The Date }
\maketitle

\begin{abstract}
We determine the behaviour of a time-varying fine structure
'constant' $\alpha (t)$ during the early and late phases of
universes dominated by the kinetic energy of changing $\alpha
(t)$, radiation, dust, curvature, and lambda, respectively. We
show that after leaving an initial vacuum-dominated phase during
which $\alpha $ increases, $\alpha $ remains constant in universes
like our own during the radiation era, and then increases slowly,
proportional to a logarithm of cosmic time, during the dust era.
If the universe becomes dominated by negative curvature or a
positive cosmological constant then $\alpha $ tends rapidly to a
constant value. The effect of an early period of de Sitter or
power-law inflation is to drive $\alpha $ to a constant value.
Various cosmological consequences of these results are discussed
with reference to recent observational studies of the value of
$\alpha $ from quasar absorption spectra and to the existence of
life in expanding universes.
\end{abstract}
\pacs{PACS Numbers: *** }
 ]

\section{Introduction}

One of the problems that cosmologists have faced in their attempts to assess
the astronomical consequences of a time variation in the fine structure
constants, $\alpha ,$ has been the absence of an exact theory describing
cosmological models in the presence of varying $\alpha .$ Until recently, it
has not been possible to analyse the behaviour of varying-$\alpha $
cosmologies in the same self-consistent way that one can explore universes
with varying $G$ using the Brans-Dicke or more general scalar-tensor
theories of gravity. However, we have recently extended the generalisation
of Maxwell's equations developed by Bekenstein so that this can be done
self-consistently. In a recent letter \cite{sbm} we reported numerical
studies of the cosmological evolution of varying-$\alpha $ cosmologies with
zero curvature, non-zero cosmological constant, and matter density matching
observations. They reveal important properties of varying-$\alpha $
cosmologies that are shared by other theories in which 'constants' vary via
the propagation of a causal scalar field obeying 2nd-order differential
equations. Their structure can be compared with that of varying speed of
light (VSL) theories developed in refs.\cite
{moffat93,am,ba,bm,barmag98,covvsl,sn} and with Kaluza-Klein like theories
in which constants like $\alpha $ vary at the same rate as the mean size of
any extra dimensions of space \cite{dims}.

Recent observations motivate the formulation and detailed investigation of
varying-$\alpha $ cosmological theories. The new observational
many-multiplet technique of Webb et al, \cite{murphy}, \cite{webb}, exploits
the extra sensitivity gained by studying relativistic transitions to
different ground states using absorption lines in quasar (QSO) spectra at
medium redshift. It has provided the first evidence that the fine structure
constant might change with cosmological time\cite{murphy,webb,webb2}. The
trend of these results is that the value of $\alpha $ was lower in the past,
with $\Delta \alpha /\alpha =-0.72\pm 0.18\times 10^{-5}$ for $z\approx
0.5-3.5.$ Other investigations have claimed preferred non-zero values of $%
\Delta \alpha $ $<0$ to best fit the cosmic microwave background (CMB) and
Big Bang Nucleosynthesis (BBN) data at $z\approx 10^3$ and $z\approx 10^{10}$
respectively\cite{avelino,bat}, but these need to be much larger than those
needed to reconcile the observations of \cite{murphy,webb,webb2}.

In this paper we present a detailed analytic and numerical study of the
behaviour of the cosmological solutions of the varying theory presented in
\cite{sbm}. We shall confine our attention to universes containing dust and
radiation but analyse the effects of negative spatial curvature and a
positive cosmological constant. Extensions to general perfect-fluid
cosmologies can easily be made if required.

\section{A simple Varying-Alpha Theory}

The idea that the charge on the electron, or the fine structure constant,
might vary in cosmological time was proposed in 1948 by Teller, \cite{tell},
who suggested that $\alpha \propto (\ln t)^{-1}$ was implied by Dirac's
proposal that $G\propto t^{-1}$ and the numerical coincidence that $\alpha
^{-1}\sim \ln (hc/Gm_{pr}^2)$, where $m_{pr\text{ }}$is the proton mass.
Later, in 1967, Gamow \cite{gam} suggested $\alpha \propto t$ as an
alternative to Dirac's time-variation of the gravitation constant, $G$, as a
solution of the large numbers coincidences problem but in 1963 Stanyukovich
had also considered varying $\alpha $, \cite{stan}, in this context. It had
the advantage of not producing a terrestrial surface temperature above $100$
degrees centigrade in the pre-Cambrian era when life was known to exist.
However, this power-law variation in the recent geological past was soon
ruled out by other evidence.

There are a number of possible theories allowing for the variation of the
fine structure constant, $\alpha $. In the simplest cases one takes $c$ and $%
\hbar $ to be constants and attributes variations in $\alpha $ to changes in
$e$ or the permittivity of free space (see \cite{am} for a discussion of the
meaning of this choice). This is done by letting $e$ take on the value of a
real scalar field which varies in space and time (for more complicated
cases, resorting to complex fields undergoing spontaneous symmetry breaking,
see the case of fast tracks discussed in \cite{covvsl}). Thus $%
e_0\rightarrow e=e_0\epsilon (x^\mu ),$ where $\epsilon $ is a dimensionless
scalar field and $e_0$ is a constant denoting the present value of $e$. This
operation implies that some well established assumptions, like charge
conservation, must give way \cite{land}. Nevertheless, the principles of
local gauge invariance and causality are maintained, as is the scale
invariance of the $\epsilon $ field (under a suitable choice of dynamics).
In addition there is no conflict with local Lorentz invariance or covariance.

With this set up in mind, the dynamics of our theory is then constructed as
follows. Since $e$ is the electromagnetic coupling, the $\epsilon $ field
couples to the gauge field as $\epsilon A_\mu $ in the Lagrangian and the
gauge transformation which leaves the action invariant is $\epsilon A_\mu
\rightarrow \epsilon A_\mu +\chi _{,\mu },$ rather than the usual $A_\mu
\rightarrow A_\mu +\chi _{,\mu }.$ The gauge-invariant electromagnetic field
tensor is therefore
\begin{equation}
F_{\mu \nu }=\frac 1\epsilon \left( (\epsilon A_\nu )_{,\mu }-(\epsilon
A_\mu )_{,\nu }\right) ,
\end{equation}
which reduces to the usual form when $\epsilon $ is constant. The
electromagnetic part of the action is still
\begin{equation}
S_{em}=-\int d^4x\sqrt{-g}F^{\mu \nu }F_{\mu \nu }.
\end{equation}
and the dynamics of the $\epsilon $ field are controlled by the kinetic term
\begin{equation}
S_\epsilon =-\frac 12\frac{\hbar c}{l^2}\int d^4x\sqrt{-g}\frac{\epsilon
_{,\mu }\epsilon ^{,\mu }}{\epsilon ^2},
\end{equation}
as in dilaton theories. Here, $l$ is the characteristic length scale of the
theory, introduced for dimensional reasons. This constant length scale gives
the scale down to which the electric field around a point charge is
accurately Coulombic. The corresponding energy scale, $\hbar c/l,$ has to
lie between a few tens of $MeV$ and Planck scale, $\sim 10^{19}GeV$ to avoid
conflict with experiment.

Our generalisation of the scalar theory proposed by Bekenstein \cite{bek2}
described in ref. \cite{sbm} includes the gravitational effects of $\psi $
and gives the field equations:
\begin{equation}
G_{\mu \nu }=8\pi G\left( T_{\mu \nu }^{matter}+T_{\mu \nu }^\psi +T_{\mu
\nu }^{em}e^{-2\psi }\right) .  \label{ein}
\end{equation}
The stress tensor of the $\psi $ field is derived from the lagrangian ${\cal %
L}_\psi =-{\frac \omega 2}\partial _\mu \psi \partial ^\mu \psi $ and the $%
\psi $ field obeys the equation of motion
\begin{equation}
\Box \psi =-\frac 2\omega e^{-2\psi }{\cal L}_{em}  \label{boxpsi}
\end{equation}
where we have defined the coupling constant $\omega = (\hbar c)/ l^2$. This
constant is of order $\sim 1$ if, as in \cite{sbm}, the energy scale is
similar to Planck scale. In order to make quantitative predictions we need
to know how much of the non-relativistic matter contributes to the RHS of
Eqn.~(\ref{boxpsi}). This is parametrised by $\zeta \equiv \rho _{em}/\rho
_{matter}$. In \cite{bek2}, $\zeta $ was estimated to be around $1\%$.
However, if we choose to model the proton as a charged shell of radius equal
to the estimated proton radius, the fraction would be lower, approximately $%
0.19\%$ . Also, the value of $\zeta $ needs to be weighted by the fraction
of matter that is non-baryonic, a point ignored in the literature \cite
{bek2,livio}. Hence, the total $\zeta $ depends strongly on the nature of
the dark matter. BBN predicts an approximate value for the baryon density of
$\Omega _B\approx 0.0125h_0^{-2}$, or $\Omega _B\approx 0.03\%$ with a
Hubble parameter of $h_0\approx 0.6$. Since we believe the total matter
density to be $\Omega _m\approx 0.3$, this would mean that only about $1/10$
of matter is baryonic and couples to changes in $e$. Thus, we should assume
values for $\zeta $ ranging from $0.02\%$ to $0.1\%$, if the cold dark
matter is allowed to have little or no electrostatic Coulomb component. If
this is not true, then $\zeta $ could have a much higher value.

We should not confuse this theory with other similar variations.
Bekenstein's theory \cite{bek2} does not take into account the stress energy
tensor of the dielectric field in Einstein's equations, and their
application to cosmology. Dilaton theories predict a global coupling between
the scalar and all other matter fields. As a result they predict variations
in other constants of nature, and also a different dynamics to all the
matter coupled to electromagnetism. An interesting application of our
approach has also recently been made to braneworld cosmology in \cite{youm}.

\section{The Cosmological Equations}

Assuming a homogeneous and isotropic Friedmann metric with expansion scale
factor $a(t)$ and curvature parameter $k$ in eqn. (\ref{ein}), we obtain the
field equations ($c\equiv 1$)
\begin{eqnarray}
\left( \frac{\dot{a}}a\right) ^2 &=&\frac{8\pi G}3\left( \rho _m\left(
1+\zeta \exp {[-2\psi ]}\right) +\rho _r\exp {[-2\psi ]}+\frac \omega 2\dot{%
\psi}^2\right)  \nonumber \\
&&-\frac k{a^2}+\frac \Lambda 3,  \label{fried1}
\end{eqnarray}
where $\Lambda $ is the cosmological constant. For the scalar field we have
the propagation equation,
\begin{equation}
\ddot{\psi}+3H\dot{\psi}=\frac 2\omega \exp {[-2\psi ]}\zeta \rho _m,
\label{psidot}
\end{equation}
where $H\equiv \dot{a}/a$ is the Hubble expansion rate$.$ The conservation
equations for the non-interacting radiation, and matter densities are
\begin{eqnarray}
\dot{\rho _m}+3H\rho _m &=&0 \\
\dot{\rho _r}+4H\rho _r &=&2\dot{\psi}\rho _r.  \label{conservation}
\end{eqnarray}
and so $\rho _m\propto a^{-3}$and $\rho _r$ $e^{-2\psi }\propto a^{-4},$
respectively. If additional non-interacting perfect fluids satisfying
equation of state $p=(\gamma -1)\rho $ are added to the universe then they
contribute density terms $\rho \propto a^{-3\gamma }$ to the RHS of eq.(\ref
{fried1}) as usual. This theory enables the cosmological consequences of
varying $e$, to be analysed self-consistently rather than by changing the
constant value of $e$ in the standard theory to another constant value, as
in the original proposals made in response to the large numbers coincidences
(see ref. \cite{bt} for a full discussion).

We have been unable to solve these equations in general except for a few
special cases. However, as with the Friedmann equation of general
relativity, it is possible to determine the overall pattern of cosmological
evolution in the presence of matter, radiation, curvature, and positive
cosmological constant by matched approximations. We shall consider the form
of the solutions to these equations when the universe is successively
dominated by the kinetic energy of the scalar field $\psi $, pressure-free
matter, radiation, negative spatial curvature, and positive cosmological
constant$.$ Our analytic expressions are checked by numerical solutions of (%
\ref{fried1}) and (\ref{psidot}).

\subsection{The Dust-dominated era}

We consider first the behaviour of dust-filled universes far from the
initial singularity. We assume that $k=0=\Lambda =\rho _\gamma ,$ so the
Friedmann equation (\ref{fried1}) reduces to

\begin{equation}
\left( \frac{\dot{a}}a\right) ^2=\frac{8\pi G}3\left( \rho _m\left( 1+\zeta
\exp{[-2\psi] }\right) +\frac \omega 2\dot{\psi}^2\ \right) ,
\label{fried1dust}
\end{equation}
and seek a self-consistent approximate solution in which the scale factor
behaves as

\begin{eqnarray}
a &=&t^{2/3}  \label{dust1} \\
\frac d{dt}(\dot{\psi}a^3) &=&N\exp{[-2\psi] }  \label{dust2}
\end{eqnarray}
where

\begin{equation}
N\equiv \frac{2\zeta }\omega \rho _ma^3  \label{dust3}
\end{equation}
is a positive constant. If we put

\[
x=\ln (t)
\]
then (\ref{dust2}) becomes
\begin{equation}
\psi ^{\prime \prime }+\psi ^{\prime }=N\exp [-2\psi ]  \label{ev}
\end{equation}
with $N\geq 0$ and $^{\prime }\equiv d/dx.$ This equation has awkward
behaviour. For any power-law behaviour of the scale factor other than (\ref
{dust1}) a simple exact solution of (\ref{dust2}) exists. However, the
late-time dust solutions are exceptional, reflecting the coupling of the
charged matter to the variations in $\psi $, and are approximated by the
following asymptotic series:

\begin{equation}
\psi = {\frac{ 1 }{2}}\ln [2Nx]+\sum_{n=1}^\infty a_nx^{-n}  \label{sol}
\end{equation}
To see this, substitute this in the evolution eqn. (\ref{ev}) for $\psi $
then it becomes:

\begin{eqnarray}
-\frac 1{2x^2}+\sum_{n=1}n(n+1)a_nx^{-n-2} & &  \nonumber \\
+\frac 1{2x} -\sum_{n=1}na_nx^{-n-1}&=&\frac 1{2x}\exp
[-2\sum_{n=1}a_nx^{-n}]  \label{sol2}
\end{eqnarray}
Now we can pick the $a_n$ to cancel out all the terms in $x^{-r}$ , $r\geq 2$
on the left-hand side. This requires

\[
a_2=a_1=-\frac 12,a_3=2a_2,a_4=3a_3=3\times 2a_2,etc
\]
hence

\begin{eqnarray}
\sum_{n=1}a_nx^{-n}&=&-\frac 12\{\frac 1x+\frac 1{x^2}+\frac 2{x^3}
\nonumber \\
& & +\frac{2\times 3}{x^4}+\frac{2\times 3\times 4}{x^5}+...+\frac{(r-1)!}{%
x^r}+..\}  \nonumber
\end{eqnarray}
all that is left of the eqn. (\ref{sol2}) is

\[
\frac 1{2x}=\frac 1{2x}\exp [-2\sum_{n=1}a_nx^{-n}]\rightarrow \frac 1{2x}%
\text{ }
\]
as $x\rightarrow \infty .$ So, at late times, as $x=\ln (t)$ becomes large,
we have

\begin{eqnarray}
\psi &=& {\frac{1 }{2}}\ln [2N(\ln (t))] -\frac 12\{\frac 1{\ln (t)}+\frac 1{%
(\ln (t))^2} +\frac 2{(\ln (t))^3}  \nonumber \\
& & +\frac{2\times 3}{(\ln (t))^4}+\frac{2\times 3\times 4}{(\ln (t))^5}+...+%
\frac{(r-1)!}{(\ln (t))^r}+..\};  \label{psi}
\end{eqnarray}
also, since $\alpha =\exp [2\psi ]$ we have, as $t\rightarrow \infty $

\begin{eqnarray}
\alpha &=&2N\ln(t)\times \exp [-\frac 1{\ln (t)}-\frac 1{(\ln (t))^2}-\frac 2%
{(\ln (t))^3}  \nonumber \\
& &-\frac{2\times 3}{(\ln (t))^4}-\frac{2\times 3\times 4}{(\ln (t))^5}-...-%
\frac{(r-1)!}{(\ln (t))^r}-].  \label{ald}
\end{eqnarray}
So, to leading order, we have

\begin{equation}
\alpha \sim 2N\ln(t)\exp [-\frac 1{\ln (t)}]  \label{ald2}
\end{equation}

\begin{center}
\begin{figure}[tbp]
\psfig{file=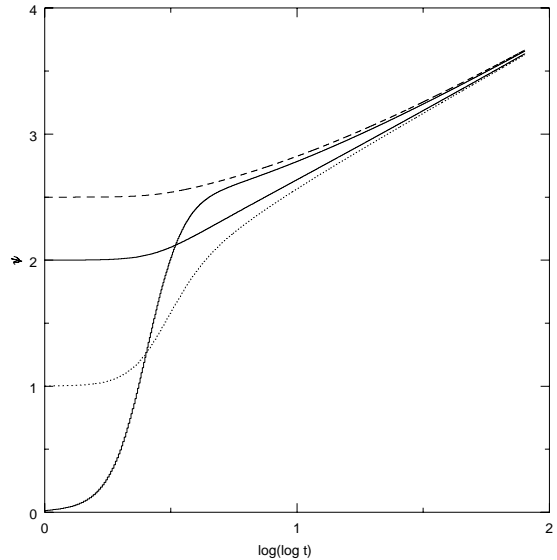,width=8cm}
\caption{Numerical solution to the equations in the dust-dominated epoch. $%
\psi $ is plotted against $log(logt)$, with initial conditions $\psi
=0,1,2,2.5$. The numerical solution clearly approaches the asymptotic
solution in the expected manner. The time is plotted in Planck units of $%
10^{-43}s.$}
\label{matfig}
\end{figure}
\end{center}

The non-analytic $\exp[1/x]$ behaviour shows why the eqn. (\ref{ev}),
despite looking simple, has awkward behaviour. We can simplify the
asymptotic series (\ref{ald}) a bit further because we know from the
definition of the logarithmic integral function $li(x)=\int_0^xdt/\ln (t)=%
{\rm Ei}[\ln (x)],$ that as $x\rightarrow \infty $

\begin{equation}
li(x)\sim \exp[x]\sum_{n=0}^\infty \frac{n!}{x^{n+1}}
\end{equation}
so the series we have in (\ref{psi}) in \{..\} brackets is

\begin{equation}
\sum_{r=1}^\infty \frac{(r-1)!}{x^r}\sim \exp{[-x]}li(\exp[x])
\end{equation}
and so asymptotically

\begin{equation}
\psi = {\frac{1 }{2}}\ln [2Nx]-\frac 12 \exp{[-x]}li(\exp[x]).
\end{equation}
Hence, as $t\rightarrow \infty ,$

\begin{equation}
\psi ={\frac 12}\ln [2N\ln (t)]-\frac 1{2t}\ li(t)={\frac 12}\ln [2N\ln (t)]-%
\frac 1{2t}{\rm Ei}[\ln (t)]  \label{pd1}
\end{equation}
and so asymptotically,

\begin{equation}
\alpha =\exp [2\psi ]=2N\exp [-t^{-1}li(t)]\ln t.  \label{pd2}
\end{equation}
This asymptotic behaviour is confirmed by solving equations (\ref{fried1}-%
\ref{conservation}) numerically for $\rho_m \gg \rho_r , \rho_\psi$. By
using a range of initial values for $\psi$ we produce the plot in fig (\ref
{matfig}), in which the asymptotic solution is clearly approached.

We need to check that the original assumption of $a=t^{2/3}$ in the
Friedmann eqn. (\ref{fried1}) is self consistent. The relevant terms are

\begin{equation}
\rho _m\left( 1+\zeta \exp {[-2\psi ]}\right) +\frac \omega 2\dot{\psi}^2
\end{equation}
The $\exp [{-2\psi }]$ $=\alpha ^{-1}$ falls off as $t\rightarrow \infty $
so the $\rho _m\left( 1+\zeta \exp {[-2\psi ]}\right) \propto a^{-3}$ term
dominates as expected. For the kinetic term $\dot{\psi}^2$ we have

\begin{equation}
\dot{\psi}=\frac 1t\times O(\frac 1{\ln (t)})
\end{equation}
and so again the $\dot{\psi}^2$ term falls off faster than $t^{-2}$ as $%
t\rightarrow \infty $ and the $a=t^{2/3}$ behaviour is an ever-improving
approximation at late times. If we examine the form of the solution (\ref
{pd2}) we see that $\alpha $ always {\it increases }with time as a
logarithmic power until it grows sufficiently for the exponential term on
the right-hand side of (\ref{psidot}) to affect the solution significantly
and slow the rate of increase by the series terms. The rate at which $\alpha
$ grows is controlled by the total density of matter in the model, which is
directly proportional to the constant $N$, defined by eqn. (\ref{dust3}).
The higher the density of matter (and hence $N$) the faster the growth in $%
\alpha $. However, because of the logarithmic time-variation, the dependence
on $\rho _m,\omega ,$ and $\zeta $ is weak. The self-consistency of the
usual $a=t^{2/3}$ dust evolution for the scale factor leaves the standard
cosmological tests unaffected. This is just as one expects for the very
variations indicated by the observations of \cite{webb2}.

%\emph{Some pictures and discussion of numerical results for the dust models
%here}

\subsection{The Radiation-dominated era}

In the radiation era we assume $k=\Lambda =0$ and take $a=t^{1/2}$ as the
leading order solution to (\ref{fried1}). We must now solve

\begin{equation}
\frac d{dt}\left( \dot{\psi}a^3\right) =N\exp [-2\psi ].  \label{sca}
\end{equation}

There is a simple particular exact solution

\begin{equation}
\psi =\frac 12\ln (8N)+\frac 14\ln (t)  \label{exrad}
\end{equation}

Consider a perturbation of this solution by $f(t)$

\begin{eqnarray*}
\psi &=&\frac 12\ln (8N)+\frac 14\ln (t)+f(t) \\
&&
\end{eqnarray*}
Inserted in eqn. (\ref{sca}) we then get %\begin{eqnarray*}
\begin{equation}
\ddot{f}+\frac 3{2t}\dot{f} =\frac 1{8t^2}(\exp{[-2f]}-1)  \label{perturbeqn}
\end{equation}
%\end{eqnarray*}

Let us first consider the case of a large perturbation, $\exp {(-2f)}\ll 1$.
The RHS of (\ref{perturbeqn}) then reduces to $-1/(8t^2)$, and through a
straightforward integration we get
\begin{equation}
\dot{f}=-\frac 1{4t}+Ct^{-3/2}
\end{equation}
with $C$ an arbitrary constant. As $t$ increases this will approach $-1/(4t)$
which has the same absolute value and is opposite in sign to the derivative
of the exact solution (\ref{exrad}). Thus for values of $\psi $ much higher
than this solution $\dot{\psi}$ is zero. $\psi $ will stay constant until
the perturbation $f$ becomes small and $\psi $ approaches the exact solution
(\ref{exrad}).

To establish the stability of the exact solution we need to consider small
perturbations around it. For small $f$ we have

\begin{equation}
\ddot{f}+\frac 3{2t}\dot{f}+\frac 1{4t^2}f=0.
\end{equation}

Hence,

\begin{equation}
f=\frac 1t\{A\sin [\sqrt{3}\ln (t)]+B\cos [\sqrt{3}\ln (t)]
\end{equation}
Thus, we have

\begin{eqnarray}
\psi &\rightarrow &\frac 12\ln (8N)+\frac 14\ln (t)  \nonumber \\
& & +\frac 1t\{A\sin [\sqrt{3}\ln (t)]+B\cos [\sqrt{3}\ln (t)]\}
\label{rad1} \\
\alpha &=&e^{2\psi }\rightarrow 8Nt^{1/2}\exp [\frac 2t\{A\sin [\sqrt{3}\ln
(t)]  \nonumber \\
& & +B\cos [\sqrt{3}\ln (t)]\}]\rightarrow 8Nt^{1/2}  \label{rad2}
\end{eqnarray}
as $t\rightarrow \infty .$

We need to check that the $\dot{\psi}^2$ term does not dominate as $%
t\rightarrow \infty$. We have

\begin{equation}
\dot{\psi}\sim \frac 1{4t}+\frac 1{t^2}\times oscillations
\end{equation}
Thus the $\dot{\psi}^2$ term is the {\em same order} of $t$ as the radiation
density term if we assume $a\sim t^{1/2}.$ Also, the matter density term $%
\rho _m\left( 1+\zeta \exp{[-2\psi] }\right) \sim \rho _m \exp{[-2\psi] }%
\sim a^{-3}\exp{[-2\psi] }\sim t^{-3/2}\times t^{-1/2}\sim t^{-2}$ is the
same order of time variation as the radiation-density term because of the
variation in $\alpha $. The assumption $a=t^{1/2}$ is still good
asymptotically but there is an algebraic constraint from the Friedmann eqn. (%
\ref{fried1})

Evaluating the terms in (\ref{fried1}), we have

\begin{equation}
\frac 1{4t^2}=\frac{8\pi G}3\left( \frac M{t^{3/2}}[1+\frac S{8Nt^{1/2}}%
]\right) +\frac \Gamma {t^2}+\frac \omega {32t^2}
\end{equation}
where $\rho _m=Ma^{-3},$ $\rho _\gamma \exp{[-2\psi] }=\Gamma a^{-4},$ $%
N=2M\zeta \omega ^{-1}$ and we have $\zeta \sim 0.02\%-0.1\%$ and probably $%
\omega \sim 1.$ So, to $O(t^{-2}),$ we have the algebraic constraint

\begin{eqnarray*}
\frac 14 &=&\frac{8\pi G}3[\frac{3\omega }{32}+\Gamma ] \\
&&
\end{eqnarray*}
This generalises the familiar general relativity $(\omega =0)$ radiation
universe case where we have $\Gamma =3/32\pi G.$

\begin{figure}[tbp]
\psfig{file=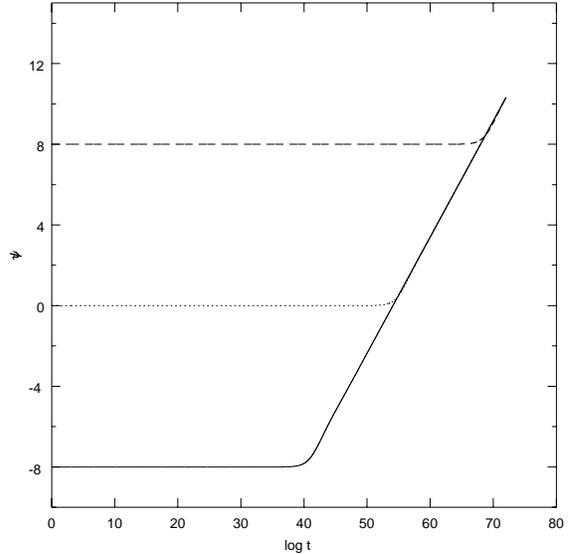,width=8cm} \caption{Numerical solution to
the equations in the radiation-dominated epoch given different
initial conditions. The particular exact solution is eventually
reached in all cases. The time is plotted in units of the Planck
time.} \label{radfig}
\end{figure}

Again, the asymptotic behaviour in eqns. (\ref{rad1})-(\ref{rad2}), and the
approach to the exact solution (\ref{exrad}), can be confirmed by numerical
solutions to eqns.(\ref{fried1}) - (\ref{conservation}) in the case of
radiation domination. The results from runs with initial values for $\psi
=-8,0,8$, $\dot{\psi}=0$ and same value for $N$, are shown in fig.(\ref
{radfig}). The particular solution (\ref{exrad}) is clearly an attractor. It
is also seen that if the system starts off with values higher than $%
1/2ln(8N) $, $\psi $ will stay constant until it reaches the value of the
solution, as predicted above.
%We see that the asymptotic solution (\ref{rad1})-(\ref{rad2}), and the exact
%solution (\ref{exrad}) it attracts to, have $\psi $ increasing as $%
%t\rightarrow \infty $. However, for $t<64N^2$ the constant term $\frac 12\ln
%(8N)$ will dominate the $\frac 14\ln (t)$ term and $\alpha $ will remain
%approximately constant.
In cosmological models containing matter and radiation with densities given
by those observed in our universe this is the case, as seen in the
computations shown in ref. \cite{sbm}. Hence, during the radiation era $%
\alpha $ remains approximately constant until the dust era begins.

%\emph{Add numerical pictures of pure radiation evolutions}
%\begin{center}
%\end{center}

This analysis can easily be extended to other equations of state. If the
Friedmann equation contains a perfect fluid with equation of state $%
p=(\gamma -1)\rho $ with $\gamma \neq 0,1,2$ then there is a late time
solution of (\ref{fried1}) and (\ref{psidot}) of the form

\begin{eqnarray}
a &=&t^{\frac 2{3\gamma }}  \label{pf1} \\
\psi &=&\frac 12\ln [\frac{N\gamma ^2}{(\gamma -1)(2-\gamma )}]+\left( \frac{%
\gamma -1}\gamma \right) \ln (t)  \label{pf2} \\
&&  \nonumber
\end{eqnarray}
which reduces to (\ref{exrad}) when $\gamma =4/3.$ This solution only exists
for fluids with $1<\gamma <2.$

\subsection{The Curvature-dominated era}

In our earlier study \cite{sbm} we showed that the evolution of $\alpha $
stops when the universe becomes dominated by the cosmological constant. This
behaviour also occurs when an open universe becomes dominated by negative
spatial curvature. In a curvature-dominated era we assume that (\ref{fried1}%
) has the Milne universe solution with

\begin{equation}
a=t.
\end{equation}
We must now solve eq. (\ref{sca}) again. It has the form

\begin{equation}
\frac d{dt}\left( \dot{\psi}t^3\right) =N\exp [-2\psi ].
\end{equation}
We seek a solution of the form

\begin{equation}
\psi =\frac 12+f(t)
\end{equation}
Hence, for small $f$

\begin{equation}
\ddot{f}+\frac 3t\dot{f}+\frac{2N}{t^2}f=0
\end{equation}
Solutions exist with $f\propto t^n$ and

\begin{equation}
n=-1\pm \sqrt{1-2N}
\end{equation}
Since $N>0$ we see that the real part of $n$ is always decaying and so

\begin{equation}
\psi \rightarrow const
\end{equation}
as $t\rightarrow \infty .$ Thus, as $t\rightarrow \infty $ we have

\begin{equation}
\alpha \sim \ \alpha _\infty \exp [2At^{-1\pm \sqrt{1-2N}}],  \label{cu}
\end{equation}
where $\alpha _\infty $ and $A$ are constants.

Again we need to check that the $\dot{\psi}^2$ term does not come to
dominate. We have $\dot{\psi}^2\sim t^{2(n-1)}$ as $t\rightarrow \infty $
and this always falls faster than $ka^{-2}\propto t^{-2}$ since $n\leq 0,$
so our approximation is always good. Thus we have shown that in open
Friedmann universes $\alpha $ rapidly approaches a constant value after the
universe becomes curvature dominated. The rate of approach is controlled by
the matter density through the constant $N$ in eq. (\ref{cu}).

\begin{figure}[tbp]
\psfig{file=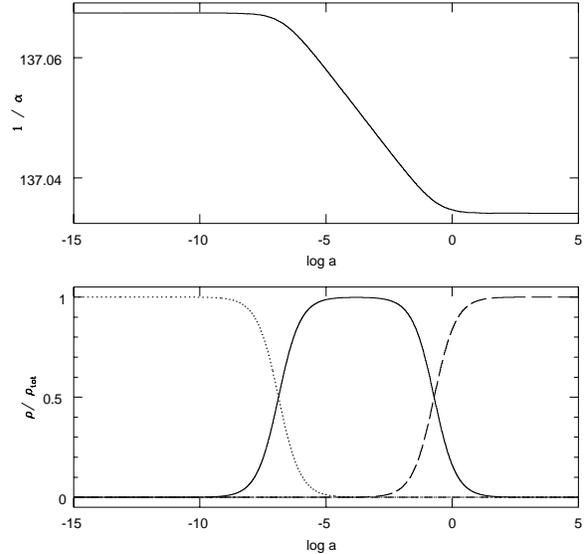,width=8cm} \caption{The top plot shows
evolution of $\alpha $ from radiation domination
through matter domination and into curvature domination where the change in $%
\alpha $ comes to an end. The lower plot shows radiation (dotted), matter
(solid) and curvature (dashed) densities as fractions of the total energy
density}
\label{curvfig}
\end{figure}

This behaviour is again confirmed by numerical solution. Fig.(\ref{curvfig})
shows how alpha changes through the dust-epoch and how the change comes to
an end as curvature takes over the expansion.

%\begin{flushright}
%\end{flushright}

\subsection{The Lambda-dominated era}

\begin{figure}[tbp]
\psfig{file=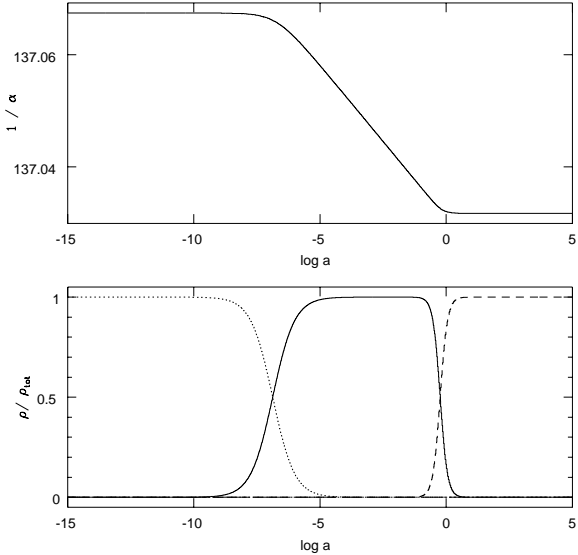,width=8cm} \caption{The top figure
shows numerical evolution of $\alpha $ from radiation domination
through matter domination and into lambda domination where the
change in $\alpha $ comes to an end. The lower plot shows
radiation (dotted), matter (solid) and lambda (dashed) densities
as fractions of the total energy density} \label{lambdafig}
\end{figure}

We can prove what was displayed in the numerical results of \cite{sbm}, and
again in fig.(\ref{lambdafig}) for the $\Lambda $-dominated era when the
value of $\Lambda $ matches that inferred from recent high redshift
supernova observations \cite{super}. At late times we assume the scale
factor to take the form

% We assume
%that at late times (\ref{fried1}) has the asymptotic solution

\begin{equation}
a =\exp [\lambda t]  \nonumber
\end{equation}
where $\lambda \equiv \sqrt{\frac \Lambda 3 }$ and so eqn.(\ref{fried1})
becomes
\begin{equation}
\frac d{dt}\left( \dot{\psi}e^{3\lambda t}\right) = N\exp [-2\psi]
\end{equation}
Linearising in $\psi $, we have

\begin{equation}
\ddot{\psi}+3\lambda \dot{\psi}=N\exp[-3\lambda t].
\end{equation}
Hence,

\begin{equation}
\psi =\psi _0+A\exp [-3\lambda t]-\frac{Nt}{3\lambda }\exp [-3\lambda
t]\rightarrow \psi _0
\end{equation}
as $t\rightarrow \infty $, where $A,\psi _0$ are arbitrary constants. Thus $%
\alpha $ approaches a constant with double-exponential rapidity during a $%
\Lambda $-dominated phase of the universe. The dominant term controlling the
late-time approach to the constant solution is proportional to the matter
density via the constant $N.$ %\emph{Numerical example?}

\subsection{Inflationary Universes}

The behaviour found for lambda-dominated universes enables us to understand
what would transpire during a period of de Sitter inflation during the early
stages of a varying-$\alpha $ cosmology. It is straightforward to extend
these conclusions to any cosmology undergoing power-law inflation. Suppose
the varying-$\alpha $ Friedmann model contains a perfect fluid with $%
p=(\gamma -1)\rho $ and $0<\gamma <2/3$. The expansion scale factor will
increase with $a(t)\propto t^{2/3\gamma },$ while $\psi $ will be governed,
to leading order by

\begin{equation}
(\dot{\psi}t^{2/\gamma }\dot{)}=0
\end{equation}

Hence, for large expansion

\begin{equation}
\psi =\psi _0+Dt^{-(2-\gamma )/\gamma }\rightarrow \psi _0
\end{equation}

and so $\psi $ and $\alpha $ approach a constant with power-law
(exponential) rapidity during any period of power-law (de Sitter) inflation.
If we evaluate the kinetic term $O(\dot{\psi}^2)$ in the Friedmann equation
and the terms $O(N\exp [-2\psi ])$ in the $\psi $ conservation equation, we
see that the assumption of $a(t)\propto t^{2/3\gamma }$ is an increasingly
good approximation as inflation proceeds. Similar behaviour would be
displayed by a quintessence field which violated the strong-energy condition
and came to dominate the expansion of the universe at late times. It would
turn off the time variation of the fine structure constant in the same
manner as the curvature of lambda terms discussed above. Note that the $\psi
$ field itself is not a possible source of inflationary behaviour in these
models. We are assuming that the inflation is contributed, as usual, by some
other scalar matter field with a self-interaction potential. However, if
this field was charged then these conclusions could be altered as the
coupling of the inflationary scalar field to the $\psi $ field would be more
complicated.

\subsection{The Very Early Universe ($t\rightarrow 0$)}

As $t\rightarrow 0$ we expect (just as in Brans-Dicke theory) to encounter a
situation where the kinetic energy of $\psi $ dominates the evolution of $%
a(t).$ This is equivalent to the solution approaching a vacuum solution of (%
\ref{fried1})-(\ref{psidot}) with $\rho _m=\rho _r=0,$ as $t\rightarrow 0$. $%
\ $In the flat case with $\Lambda =0$ (the $k\neq 0$ and $\Lambda \neq 0$
cases can be solved straightforwardly and the models with $\rho _r\neq 0$
can also be solved exactly in parametric form.) we have

\begin{equation}
\left( \frac{\dot{a}}a\right) ^2=\frac{4\pi G\omega }3\dot{\psi}^2
\end{equation}

\begin{equation}
\ddot{\psi}+3H\dot{\psi}=0
\end{equation}
Thus the exact vacuum solution is

\begin{eqnarray}
\psi &=&\psi _0+\frac 1{\sqrt{12\pi G\omega }}\ln (t) \\
a &=&t^{1/3}
\end{eqnarray}
During this phase the fine structure constant {\it increases} as a power-law
of the comoving proper time as $t$ increases:

\begin{equation}
\alpha =\exp [2\psi ]\propto t^{\frac 1{\sqrt{3\pi G\omega }}}
\end{equation}

Note that the matter and radiation density terms fall off slower than $\dot{%
\psi}^2\propto t^{-2}$ as $t\rightarrow 0$ and $\exp [-2\psi ]\propto t^{-1
/ (\sqrt{3\pi G\omega})}.$ They will eventually dominate the evolution at
some later time and the vacuum approximation will break down. As in
Brans-Dicke cosmology \cite{nar} we expect the general solutions of the
cosmological equations to approach this vacuum solution as $t\rightarrow 0$
and to approach the other late-time asymptotes discussed above as $%
t\rightarrow \infty .$

\section{Discussion}

The overall pattern of cosmological evolution is clear from the results of
the last section even though it is not possible to solve the Friedmann
equation exactly in most cases. There are five distinct phases:

\begin{itemize}
\item  {\bf a}. Near the initial singularity the kinetic part of scalar
field $\psi $ will dominate the expansion and the universe behaves like a
general relativistic Friedmann universe containing a massless scalar or
stiff perfect fluid field, with $a=t^{1/3}.$ During this 'vacuum phase', the
fine structure constant increases as a power law in time.

{\bf b}. As the universe ages the radiation density will eventually become
larger than the kinetic energy of the $\psi $ field. In this radiation
dominated epoch, the fine structure constant will approach a specific
solution, $\alpha \propto t^{1/2}$ asymptotically. In reality however, if
the initial value of $\alpha $ is much larger than the specific solution, we
will have a potentially very long transient period of constant evolution,
and the universe may become dust dominated while $\alpha $ is still constant.

{\bf c}. After dust domination begins, $\alpha $ slowly approaches an
asymptotic solution, $\alpha =2N\ln (t)\times \exp [-t^{-1}li(t)]$, where $%
li(t)$ is the logarithmic integral function. If the universe has zero
curvature and no cosmological constant this will approach the late time
solution $\alpha \propto \ln (t)$.

{\bf d}. If the universe is open then this increase will be brought to an
end when the universe becomes dominated by spatial curvature and $\alpha $
will approach a constant. If the curvature is positive the universe will
eventually reach an expansion maximum and contract so long as there are no
fluids present which violate the strong energy condition. The behaviour of
closed universes also offers a good approximation to the evolution of bound
spherically symmetric density inhomogeneities of large scale in a background
universes and will be discussed in a separate paper.

{\bf e}. If there is a positive cosmological constant, the change in $\alpha
$ will be halted when the cosmological constant starts to accelerate the
universe. If any other quintessential perfect fluid with equation of state
satisfying $p<-\rho /3$ is present in the universe then it would also
ultimately halt the change in $\alpha $ when it began to dominate the
expansion of the universe.
\end{itemize}

To obtain a more holistic picture of the evolution it is useful to string
these different parts together. To a good approximation we know that in the
vacuum phase from the Planck time $t_p$ until $t_v$ we have
\begin{equation}
a\propto t^{\frac 13};\alpha \propto t^A;A=\frac 1{\sqrt{3\pi G\omega }}
\end{equation}
In the radiation era we have $\alpha $ constant until the growth kicks in at
a time $t_{growth}$. The fine structure constant then increases as
\begin{equation}
a\propto \alpha \propto t^{1/2}
\end{equation}
until $t_{eq\text{ }}$when the radiation era end and dust takes over.
However, in universes like our own, this growth era is never reached. Then,
in the dust era,
\begin{equation}
\alpha \propto \ln t
\end{equation}
until the curvature or lambda eras begin at $t_c$ or $t_\Lambda $, after
which $\alpha $ remains constant until the present , $t_0$. So, matching
these phases of evolution together we can express $\alpha (t_0)$ in terms of
$\alpha (t_p):$

When the universe is open with $\Lambda =0:$
\begin{equation}
\alpha (t_0)=\alpha (t_p)\left( \frac{t_v}{t_p}\right) ^A\left( \frac{t_{eq}%
}{t_{growth}}\right) ^{1/2}\left( \frac{\ln (t_c/t_p)}{\ln (t_{eq}/t_p)}%
\right) ,
\end{equation}
where we have used the fact that our log formula to express ages in Planck
time units.

When the universe is flat with $\Lambda >0:$
\begin{equation}
\alpha (t_0)=\alpha (t_p)\left( \frac{t_p}{t_v}\right) ^A\left( \frac{t_{eq}%
}{t_{growth}}\right) ^{1/2}\left( \frac{\ln (t_\Lambda /t_p)}{\ln
(t_{eq}/t_p)}\right)
\end{equation}
and $t_c$ has been replaced by $t_\Lambda $.

For the radiation era we consider two extreme cases. We look at a constant $%
\alpha $ scenario with $t_{growth}=t_{eq}$ and a scenario where it grows
throughout the radiation era, $t_{growth}=t_v$.

Typically, $t_c/t_p\sim t_\Lambda /t_p\sim 10^{59}$ and $t_{eq}/t_p\sim
10^{53},$ so in both cases for constant $\alpha $ evolution in the radiation
epoch we get
\begin{equation}
\alpha (t_0)=\alpha (t_p)\left( \frac{t_v}{t_p}\right) ^A\left( \frac{59}{53}%
\right) \sim 1.11\alpha (t_p)\left( \frac{t_v}{t_p}\right) ^A
\end{equation}
% (59/53)^2 = 1.24
We approximate the value for $t_v\sim t_p\sim 1$, so for continuous growth
through radiation epoch we get
\begin{equation}
\alpha (t_0)=\alpha (t_p)\left( \frac{t_v}{t_p}\right) ^A\left(
10^{53}\right) ^{1/2}\left( \frac{59}{53}\right) \sim 10^{26}\alpha
(t_p)\left( \frac{t_v}{t_p}\right) ^A
\end{equation}
Hence there are very different possibilities for the change in $\alpha $
depending on the evolution in the radiation era.
%In general the $1.24$ depletion factor from the dust-era evolution
%will be
%dominated by any very short period of kinetic evolution during which time $%
%\alpha $ grows as a power-law

We have proved this sequence of phases by an exhaustive numerical and
analytical study. The ensuing scenario finds two interesting applications,
with which we conclude.

In \cite{sbm} we found that our theory could fit simultaneously the varying $%
\alpha $ results reported in \cite{webb,murphy,webb2} and the evidence for
an accelerating universe presented in \cite{super}. We noted the curious
fact that there is a coincidence between the redshift at which the universe
starts accelerating and the redshift where variations in $\alpha $ have been
observed but below which $\alpha $ must stabilise to be in accord with
geochemical evidence \cite{sh,fujii}. This may be explained dynamically in
our theory by the fact that the onset of lambda domination suppresses
variations in $\alpha $. Therefore $\alpha $ remains almost constant in the
radiation era, undergoes small logarithmic time-increase in the matter era,
but approaches a constant value when the universe starts accelerating
because of the presence of a positive cosmological constant. Hence, we
comply with geological, nucleosynthesis, and microwave background radiation
constraints on time-variations in $\alpha $, while fitting simultaneously
the observed accelerating universe and the recent high-redshift evidence for
small $\alpha $ variations in quasar spectra.

We have also noted that within this theory the usual anthropic arguments for
a lambda free universe may be reversed \cite{bms1}. Usually, the anthropic
principle is used to justify the near flatness and $\Lambda \approx 0$
nature of our universe since large curvature and lambda prevents the
formation of galaxies and stars from small perturbations. We have shown that
it might be anthropically {\it disadvantageous} for a universe to lie too
close to flatness or for the cosmological constant to lie too close to zero.
This constraint occurs because ``constants'' change throughout the
dust-dominated period when the curvature and lambda do not influence the
expansion of the universe. . The onset of a period of lambda or curvature
domination has the property of dynamically stabilising the constants,
thereby creating favourable conditions for the emergence of structures. If
the universe were exactly flat and lambda were exactly zero then $\alpha $
would continue to grow to a value that appears to make living complexity
impossible \cite{bms1}.

\end{document}